\documentclass[conference]{IEEEtran}
\IEEEoverridecommandlockouts
\usepackage{cite}
\usepackage{amsmath,amssymb,amsfonts}
\usepackage{graphicx}
\usepackage{textcomp}
\usepackage{subcaption} 
\usepackage{xcolor}
\usepackage{tikz}
\usetikzlibrary{arrows.meta, positioning, shapes.geometric, fit}
\usepackage[dvipsnames]{xcolor}
\usepackage[ruled,vlined,linesnumbered]{algorithm2e}

\begin{document}

\title{
CD-MED: Cross-Domain Multimodal Emotion Descriptor for Visual Comparison of Digital Objects
% A Novel Descriptor for Multi-modal Emotion Representation \\
}

\author{
\IEEEauthorblockN{Elnara Kadyrgali}
\IEEEauthorblockA{
\textit{School of Information Technology} \\ 
\textit{and Engineering} \\
\textit{Kazakh-British Technical University}\\
Almaty, Kazakhstan \\
https://orcid.org/0009-0000-4732-2806}
\and
\IEEEauthorblockN{Muragul Muratbekova}
\IEEEauthorblockA{
\textit{School of Information Technology} \\ 
\textit{and Engineering} \\
\textit{Kazakh-British Technical University}\\
Almaty, Kazakhstan \\
https://orcid.org/0009-0000-9162-2945}
\and
\IEEEauthorblockN{Pakizar Shamoi}
\IEEEauthorblockA{
\textit{School of Information Technology} \\ 
\textit{and Engineering} \\
\textit{Kazakh-British Technical University}\\
Almaty, Kazakhstan \\
https://orcid.org/0000-0001-9682-0203}
}
\maketitle

\begin{abstract}
%можно перенести в интро
% \colorbox{LimeGreen}{1.Title (multimodal, modality ind-t}
% \colorbox{LimeGreen}{2.Abstract}
% We introduce a modality-independent emotional descriptor that represents cultural artifacts as distributions of fuzzy emotional memberships in a common emotional space, enabling direct comparison across heterogeneous modalities such as music, text, images, and video.
Digital objects express emotions through different modalities. For example, a movie may include visual scenes, audio, dialogue, and facial expressions, while a song may contain melody, rhythm, lyrics, and vocal tone. Because existing emotion recognition models are usually modality-specific, it is difficult to compare such objects directly. This paper proposes CD-MED, a Cross-Domain Multimodal Emotion Descriptor for representing heterogeneous digital objects in a common emotional space. Each modality can be processed by its own emotion recognition model, and the resulting emotional outputs are transformed into a shared descriptor. The descriptor preserves information from individual modalities while also allowing an integrated emotional profile of the object. For interpretation, CD-MED is visualized in the valence–arousal space: position represents affective coordinates, color denotes emotion category, size indicates intensity, and shape shows the modality. This unified representation enables emotion-based comparison, retrieval, recommendation, and visualization across different domains such as movies, songs, images, and books.

\end{abstract}


\section{Introduction}
% Emotion plays a fundamental role in human perception, communication, and decision-making \cite{bookAffectiveComputing, AffectiveComputingAndSentiment}. Across different forms of media, including images, text, speech, and video, emotional content significantly influences how humans interpret and experience content \cite{Baveye2018}. Emotional responses serve as an essential bridge between raw sensory input and subjective understanding, shaping preferences, behaviors, and interactions \cite{Eliane2016}. \colorbox{LimeGreen}{3.Add 1-2 citations}
% \colorbox{LimeGreen}{4.Fig for hypothesis}
% \colorbox{LimeGreen}{5.2-3 pars add transitions, remove fuzzy}

% Artificial Intelligence (AI) has increasingly moved beyond functional tasks toward understanding and interacting with human emotions. As a result, 

Affective Computing has emerged as a multidisciplinary field that integrates computer science, psychology, and machine learning to enable emotion-aware intelligent systems \cite{bookAffectiveComputing}. The integration of emotion recognition has become essential for modern applications, from personalized recommendation systems to digital healthcare \cite{Calvo, AlAzani2025}.

% To enable consistent emotion analysis across diverse modalities, affective computing requires a common representation framework that captures affective information in a modality-independent manner. 

Among the existing approaches, continuous dimensional models have become the dominant paradigm.
% Human emotions often exhibit fuzzy boundaries, gradual transitions, and mixed affective states. To better capture this complexity, affective computing has increasingly adopted continuous dimensional representations.  
In particular, the two-dimensional valence–arousal space has emerged as a reliable and experimentally validated framework for the quantitative analysis and representation of affective states \cite{Russell1980}. Its modality-independent nature makes it particularly well-suited to modeling emotions across heterogeneous media.

Despite these advances, existing approaches remain largely modality-specific, limiting their ability to compare or integrate emotional representations across different data types. This poses a fundamental challenge to building universal, emotion-aware systems capable of understanding and aligning affective signals in a unified way.

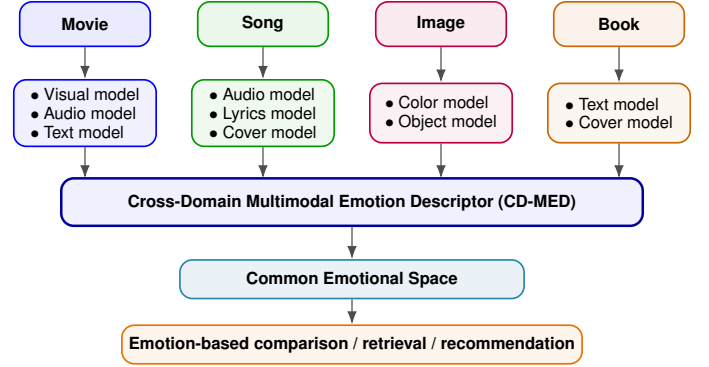
\begin{figure}[t]
\centering
\resizebox{0.5\textwidth}{!}{%
\begin{tikzpicture}[
    font=\sffamily\footnotesize,
    box/.style={
        rectangle,
        rounded corners=6pt,
        draw=#1,
        fill=#1!8,
        thick,
        minimum width=2.2cm,
        minimum height=0.75cm,
        align=center
    },
    modelbox/.style={
        rectangle,
        rounded corners=6pt,
        draw=#1,
        fill=#1!6,
        thick,
        minimum width=2.4cm,
        minimum height=1.0cm,
        align=left,
        inner xsep=6pt,
        inner ysep=4pt
    },
    mainbox/.style={
        rectangle,
        rounded corners=6pt,
        draw=blue!60!black,
        fill=blue!6,
        very thick,
        minimum width=9.8cm,
        minimum height=0.8cm,
        align=center,
        font=\sffamily\bfseries\footnotesize
    },
    smallbox/.style={
        rectangle,
        rounded corners=6pt,
        draw=#1,
        fill=#1!8,
        thick,
        minimum width=5.8cm,
        minimum height=0.65cm,
        align=center,
        font=\sffamily\bfseries\footnotesize
    },
    arrow/.style={
        -{Latex[length=2.2mm]},
        semithick,
        draw=black!70
    }
]

% Top object boxes
\node[box=blue] (movie) at (-4.5,3.3) {\textbf{Movie}};
\node[box=green!60!black] (song) at (-1.5,3.3) {\textbf{Song}};
\node[box=purple] (image) at (1.5,3.3) {\textbf{Image}};
\node[box=orange!80!black] (book) at (4.5,3.3) {\textbf{Book}};

% Model boxes
\node[modelbox=blue] (movieModel) at (-4.5,1.8)
{
$\bullet$ Visual model\\
$\bullet$ Audio model\\
$\bullet$ Text model
};

\node[modelbox=green!60!black] (songModel) at (-1.5,1.8)
{
$\bullet$ Audio model\\
$\bullet$ Lyrics model\\
$\bullet$ Cover model
};

\node[modelbox=purple] (imageModel) at (1.5,1.8)
{
$\bullet$ Color model\\
$\bullet$ Object model
};

\node[modelbox=orange!80!black] (bookModel) at (4.5,1.8)
{
$\bullet$ Text model\\
$\bullet$ Cover model
};

% Main descriptor box
\node[mainbox] (vmed) at (0,0.3)
{Cross-Domain Multimodal Emotion Descriptor (CD-MED)};

% Common space and application boxes
\node[smallbox=cyan!60!black] (space) at (0,-1.0)
{Common Emotional Space};

\node[smallbox=orange!90!black] (app) at (0,-2.1)
{Emotion-based comparison / retrieval / recommendation};

% Arrows from objects to models
\draw[arrow] (movie) -- (movieModel);
\draw[arrow] (song) -- (songModel);
\draw[arrow] (image) -- (imageModel);
\draw[arrow] (book) -- (bookModel);

% Arrows from models to CD-MED
% \draw[arrow] (movieModel.south) -- (vmed.north west);
% \draw[arrow] (songModel.south) -- (vmed.north west);
% \draw[arrow] (imageModel.south) -- (vmed.north east);
% \draw[arrow] (bookModel.south) -- (vmed.north east);
% Straight vertical arrows from models to CD-MED
\draw[arrow] (movieModel.south) -- (-4.5,0.7);
\draw[arrow] (songModel.south) -- (-1.5,0.7);
\draw[arrow] (imageModel.south) -- (1.5,0.7);
\draw[arrow] (bookModel.south) -- (4.5,0.7);
% Downward arrows
\draw[arrow] (vmed) -- (space);
\draw[arrow] (space) -- (app);

\end{tikzpicture}
}
\caption{Conceptual overview of the proposed CD-MED. Heterogeneous digital objects, e.g., movies, songs, images, and books, contain different modalities and require different modality-specific emotion recognition models. CD-MED transforms these heterogeneous emotional outputs into a common emotional space. It enables emotion-based comparison, retrieval, recommendation, and visualization across domains.}
\label{fig:vmed_concept}
\end{figure}

To address this limitation, we propose the Cross-Domain Multimodal Emotion Descriptor, CD-MED. As shown in Fig.~\ref{fig:vmed_concept}, heterogeneous digital objects may contain different modalities and require different modality-specific emotion recognition models. The proposed CD-MED maps these outputs into a common emotional space, enabling emotion-based visual comparison across domains. CD-MED is an interpretable visual representation of multimodal emotional information, in which emotion category, intensity, affective position, and source modality are encoded by color, size, position, and shape, respectively. This representation allows users to understand complex multimodal emotional profiles without directly interpreting raw model outputs.

The main contributions of this study are as follows:
\begin{itemize}
    \item We introduce the Cross-Domain Multimodal Emotion Descriptor, CD-MED, an interpretable visualization of multimodal emotional information, enabling users to understand emotion distributions across different channels through a unified visual descriptor.
    \item We propose a modality-agnostic comparison metric to measure the similarity between any two CD-MED profiles, even when they contain a different number of active modalities.
    \item We illustrate the effectiveness of CD-MED through a case study with various multimedia objects.
\end{itemize}

The structure of the paper is as follows. Section I is an introduction. Section II reviews related work on affective computing and emotion representation. Section III presents the proposed methodology. Section IV demonstrates the applicability of the proposed framework through a multimedia case study comparing movies and music. Finally, Section V concludes the paper and outlines directions for future research.

\section{Related Work}

Research in affective computing has evolved over the last few decades to enable machines to recognize, interpret, and respond to human emotions. The term \textit{Affective Computing} was first introduced by Picard \cite{bookAffectiveComputing} to describe emotion-aware computational systems. Since then, emotion analysis has become an important research area in a number of domains including computer vision, natural language processing, speech analysis, and multimedia understanding.

% \subsection{Emotion Representation Models}
Emotion representation is commonly based on two main paradigms: categorical and dimensional models \cite{cat_dim}. Categorical models describe emotions as a finite set of classes, such as happiness, sadness, fear, anger, etc. \cite{TellegenWatsonClark1999}. These models are intuitive and easy to interpret. However, they may not capture mixed or gradual emotional states. Dimensional models, such as Russell’s circumplex model \cite{Russell1980} and the valence–arousal space, represent emotions as continuous affective coordinates \cite{9881519}. Further extensions added the third dimension of dominance to give the Pleasure-Arousal-Dominance (PAD) model \cite{Mehrabian1996}. Such models are more flexible for comparing emotions across text, speech, faces, music, video, brain signals, and multimodal input because they provide a shared space for affective representation \cite{Lee2022Chinese, Praveen2022AudioVisual}.

% Early approaches to emotion representation relied on discrete emotion theories, where emotions were categorized into a finite set of classes such as happiness, sadness, anger, fear, and surprise \cite{Hevner1936 ,WatsonTellegen1985, TellegenWatsonClark1999}. 

% \subsection{Emotion Recognition in Different Modalities}
Emotion recognition has been widely studied in individual modalities, including text, speech, music, images, and video. Text-based approaches usually rely on lexicons or transformer-based language models \cite{Nandwani2021, 10409495}. Audio-based methods extract acoustic or spectral features such as pitch, rhythm, energy, and Mel-spectrograms \cite{Bhangale2023}. Image and video-based approaches analyze visual features, colors, objects, scenes, or facial expressions \cite{Zhou2023Emotion, Luo2025CVRSF-Net:}. These methods can achieve strong performance within a specific modality, but their outputs are usually not directly comparable because each modality uses different features, models, and emotion taxonomies.

% \subsection{Multimodal Affective Computing and Fusion}
Multimodal affective computing combines emotional signals from several sources, such as text, speech, images, and video \cite{Lian2023A}. Existing fusion strategies are commonly divided into feature-level (early fusion) and decision-level fusion (late fusion) \cite{Pereira2023A, poria2017}. Feature-level fusion combines modality-specific features before classification, whereas decision-level fusion combines the outputs of independent models \cite{Devi2023Multimodal, Abdullakutty2024Decoding}. These approaches improve emotion recognition accuracy when several modalities are available \cite{Wu2025A}, but they are usually designed for a specific task, dataset, or object type \cite{Pena2023A}. 

The problem of comparing objects with different sets of modalities, such as a movie and a song or a book and an image, has received limited attention in the literature \cite{Han2024Cross-Modal, Wankhade2022}. Several studies have explored shared latent spaces for cross-modal understanding, in which heterogeneous data are projected into a common embedding space to improve semantic alignment \cite{8269806}. However, most of these approaches focus on task-specific optimization rather than universal emotional representation. As a result, comparing emotional states across different media types remains a challenging problem.

% Recent advances in multimodal affective computing have focused on combining emotional signals from heterogeneous sources such as text, speech, images, and video. 

% Multimodal fusion techniques are categorized into feature-level and decision-level fusion broadly \cite{poria2017}. Feature-level fusion combines modality-specific representations before classification, whereas decision-level fusion combines independent predictions from each modality. These methods improve recognition accuracy but are still strongly task-dependent and often require aligned modality availability.

% \colorbox{LimeGreen}{6.Add citations, multimodal, fusion, synonyms of descr}
% Emotion Literature Review

% While these categorical models are intuitive and human-interpretable, they often fail to capture subtle emotional transitions and mixed affective states.

% To overcome these limitations, dimensional models of emotion were introduced, including the Russell’s circumplex model \cite{Russell1980}, which has become one of the most widely adopted frameworks. 

 % These continuous representations have demonstrated good applicability across multiple modalities.

% In contrast to existing methods, our proposed Universal Emotional Descriptor (UED) framework introduces a modality-independent emotional latent space designed for unified representation and direct comparison of heterogeneous emotional content. This allows emotional similarity to be measured regardless of the underlying media source, addressing an important limitation in current multimodal affective computing research.

\section{Methodology}
% \colorbox{Lavender}{Methodology figure - Elnara}
% \colorbox{Lavender}{7.Detailed interpretation of the methodology process - Elnara}
\subsection{Proposed Approach}

\begin{figure*}
    \centering
    \includegraphics[width=1\linewidth]{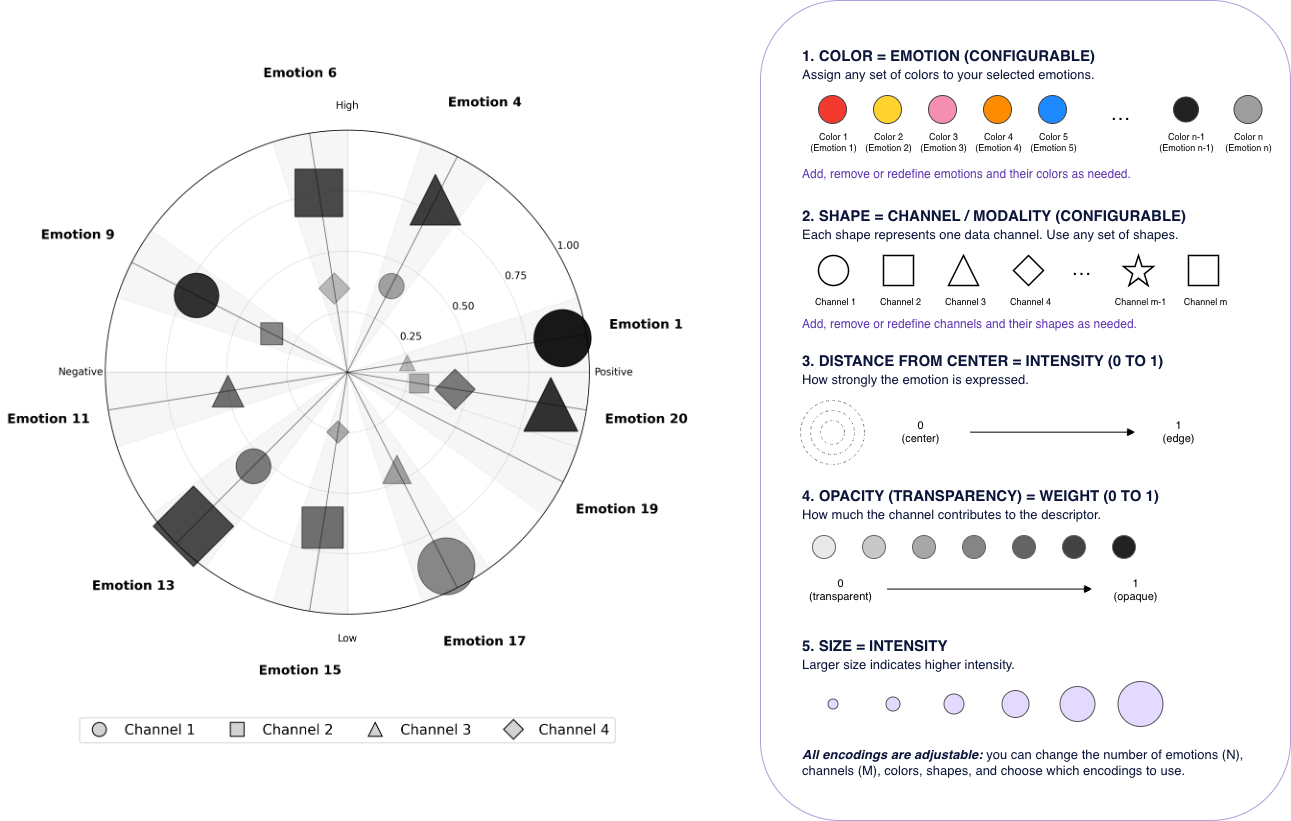}
    \caption{Prototype of the proposed emotional descriptor and its visual encoding scheme}
    \label{fig:prototype}
\end{figure*}

% формулировка нашего подхода
The proposed approach aims to represent heterogeneous multimedia artifacts in a common visual-emotional space. While text, music, images, and videos differ substantially in their low-level features, they all convey emotional information that can be represented in a common space. 

% We therefore introduce a modality-independent emotional descriptor that maps an arbitrary artifact to a fixed set of emotional dimensions. 
% Instead of comparing artifacts in their original feature spaces, the proposed framework compares their emotional representations, enabling direct analysis across different modalities.

% \subsubsection{Universal Emotional Descriptor}

Let $E={e_1,e_2,\ldots,e_k}$ denote a predefined set of emotional categories, where each dimension corresponds to a specific emotion (e.g., happiness, sadness, fear, anger, love, trust, surprise, and others). Given an arbitrary artifact $x$, its emotional representation is defined as

\begin{equation}
\mathbf{v}(x)=\left(v_1,v_2,\ldots,v_n\right),
\end{equation}

where $v_i \in [0,1]$ denotes the normalized value associated with emotion $e_i$. 

% The emotional encoding process can be formally represented as
% \begin{equation}
% f \rightarrow [0,1]^n,
% \end{equation}

% where 

For multimodal artifacts, emotional information is extracted independently from each available modality.

% Let $X = {x_1,x_2,\ldots,x_n}$ denote the set of all artifacts,each represented within the same emotional space.

% $f$ maps an artifact into the universal emotional space. Thus,

% \begin{equation}
% f(x)=\mathbf{v}(x).
% \end{equation}

% The resulting descriptor provides a modality-independent representation that enables direct comparison of emotionally similar artifacts originating from different domains, such as paintings, songs, texts, and multimodal content.

% \subsubsection{Multimodal Emotional Representation}

% This approach preserves modality-specific emotional characteristics while enabling a unified representation of complex multimedia content.

% \subsubsection{Descriptor Matrix D}

The modality-specific emotional descriptors are arranged into a descriptor matrix

\begin{equation}
\mathcal{D}
=
\begin{bmatrix}
v_{1,1} & v_{1,2} & \cdots & v_{1,n} \\
v_{2,1} & v_{2,2} & \cdots & v_{2,n} \\
\vdots & \vdots & \ddots & \vdots \\
v_{K,1} & v_{K,2} & \cdots & v_{K,n}
\end{bmatrix}
\end{equation}

where each row corresponds to a modality-specific emotional descriptor and each column corresponds to an emotional category. 

% Thus, the matrix ($\mathbf{D}$) preserves the emotional characteristics extracted from individual modalities while maintaining a common emotional representation across all channels.

% \subsubsection{Integrated Descriptor V}
% To obtain a compact representation of the multimodal artifact, the modality-specific descriptors are aggregated into a single integrated emotional descriptor

% \begin{equation}
% \mathbf{V}
% =
% \sum_{k=1}^{m}
% w_k \mathbf{v}_k,
% \end{equation}

% where ($w_k$) denotes the contribution weight assigned to modality (k) and satisfies

% \begin{equation}
% \sum_{k=1}^{m}
% w_k
% =
% 1,
% \end{equation}

The resulting descriptor matrix ($\mathcal{D}$) represents the artifact's overall emotional profile in the universal emotional space. Consequently, heterogeneous objects, including images, musical compositions, textual documents, videos, and other multimodal content, can be directly compared using a common emotional descriptor regardless of their original modality.

% \colorbox{LimeGreen}{8. Check information relevance}
% \subsection{Emotion Representation}
% наши значения Valence Arousal для каждоой эмоции - в таблице укажем
% цвета каждой эмоции
% модальность определяется фигурой (text - O, poster - square)

To provide an interpretable visualization of the proposed universal emotional descriptor, each emotional category is mapped into a two-dimensional affective space using predefined valence-arousal coordinate areas. This representation allows emotional descriptors to be analyzed not only numerically but also geometrically. Each emotion \(E_i\) is associated with a pair of affective values: the valence dimension (negative to positive affect), and the arousal dimension (low to high activation). These values are predefined based on an established psychological emotion model and summarized in Table~\ref{tab:emotion_va}.

The angular positions were derived from the original Geneva Emotion Wheel (GEW) \cite{Scherer2005, Scherer2013} representation, which contains 20 emotions distributed across a $360^\circ$ affective space. Accordingly, each emotion was assigned an angular sector of $18^\circ$ ($360^\circ / 20$). The values reported in Table~\ref{tab:emotion_va} correspond to the central angle (median) of each emotional sector. For example, the Happiness sector spans from $18^\circ$ to $36^\circ$, its representative angle is $27^\circ$.

\begin{table}[t]
\centering
\caption{Emotion angles in the valence--arousal space and color associations}
\label{tab:emotion_va}
\begin{tabular}{|l|c|l|l|}
\hline
\textbf{Emotion} & \textbf{Angle in VA} & \textbf{Proof of concept} & \textbf{Color} \\ \hline

% Happiness & $7.8^\circ$ & Russell \cite{Russell1980} 
% & Yellow \colorbox{yellow}{\phantom{XX}} \cite{Jonauskaite2025} \\ \hline

Happiness & $27^\circ$ & \cite{Scherer2005, Scherer2013}  
& \colorbox{yellow}{\phantom{XX}} Yellow\cite{Jonauskaite2025} \\ \hline

Love & $351^\circ$ & \cite{Scherer2005, Scherer2013} 
& \colorbox{pink}{\phantom{XX}} Pink \cite{Jonauskaite2025} \\ \hline

Anger & $99^\circ$ & \cite{Scherer2005, Scherer2013} 
& \colorbox{red}{\phantom{XX}} Red \cite{Jonauskaite2025} \\ \hline

% Sadness & $207.5^\circ$ & Russell \cite{Russell1980} 
% & Blue \colorbox{blue}{\phantom{XX}} \cite{Jonauskaite2025} \\ \hline

Sadness & $243^\circ$ & \cite{Scherer2005, Scherer2013} 
& \colorbox{cyan}{\phantom{XX}} Blue \cite{Jonauskaite2025} \\ \hline

% Gratitude & $20^\circ$ & Scherer \cite{Scherer2005, Scherer2013} 
% & Green \colorbox{green}{\phantom{XX}} \cite{} \\ \hline

Fear & $225^\circ$ & \cite{Scherer2005, Scherer2013} 
& \colorbox{black}{\phantom{XX}} Black \cite{Jonauskaite2025} \\ \hline

Shame & $207^\circ$ & \cite{Scherer2005, Scherer2013} 
& \colorbox{gray}{\phantom{XX}} Gray \cite{Jonauskaite2025} \\ \hline

Surprise & $297^\circ$ & \cite{Scherer2005, Scherer2013} 
& \colorbox{orange}{\phantom{XX}} Orange \cite{Jonauskaite2025} \\ \hline

% Shyness & $310^\circ$ & Scherer \cite{Scherer2005} 
% & Violet \colorbox{violet}{\phantom{XX}} \cite{} \\ \hline

% Trust & $350^\circ$ & Plutchik \cite{Plutchik1980} 
% & Blue \colorbox{blue}{\phantom{XX}} \cite{} \\ \hline

\end{tabular}
\end{table}

% Each emotion is assigned a fixed position in the emotional space, while visual attributes encode complementary information: color represents emotion category, shape denotes modality (channel), distance from the center indicates emotional intensity, opacity reflects channel weight, and marker size represents confidence or intensity.

In addition to their affective coordinates, each emotional category is assigned a unique color for visual distinction. This color encoding provides an intuitive way to identify emotional regions in the shared affective space.

To preserve modality information during visualization, each artifact is represented using a modality-specific geometric marker. For example:

\begin{itemize}
    \item Circle --- text artifacts
    \item Square --- image/poster artifacts
    \item Triangle --- music/audio artifacts
    \item Diamond --- video artifacts
\end{itemize}

Thus, while all artifacts share the same emotional coordinate system, their original modality remains distinguishable through shape encoding.

% The emotional representation of an artifact \(x\) is visualized by projecting its dominant or weighted emotional components into the valence-arousal plane. The contribution of each emotion is determined by its normalized emotional score \(v_i(x)\), allowing mixed emotional states to be represented as weighted distributions rather than single-point labels.

% The affective centroid of an artifact can be computed as:

% \begin{equation}
%     C(x) = \left( 
%     \frac{\sum_{i=1}^{n} v_i(x)\cdot Val_i}{\sum_{i=1}^{n} v_i(x)},
%     \frac{\sum_{i=1}^{n} v_i(x)\cdot Ar_i}{\sum_{i=1}^{n} v_i(x)}
%     \right).
% \end{equation}

% This centroid represents the overall emotional position of the artifact in the valence-arousal space.

Fig. \ref{fig:prototype} illustrates the prototype of the proposed multimodal emotional descriptor. Thus, the proposed emotional representation integrates the following information:

\begin{enumerate}
    \item Color - emotional category identity;
    \item Shape - original artifact modality;
    \item Position - affective location in valence-arousal space;
    \item Color Opacity - the channel weight;
    \item Size - the emotional intensity.
\end{enumerate}

In the current implementation, channel weights are not used, and all modalities contribute equally to the final representation. Nevertheless, the weighting mechanism is retained as an optional, adaptable component for application-specific scenarios that require different contributions from different modalities.

Taken together, this multimodal visualization provides an interpretable framework for exploring emotional relationships among heterogeneous artifacts in a unified emotional space.

\subsection{Comparison Strategy}
% \colorbox{LimeGreen}{Describe 3 metrics}

% Once represented in the universal emotional space, heterogeneous artifacts can be compared independently of their original modality. Since all artifacts are described by emotional descriptors of identical structure, similarity can be evaluated directly in the emotional space.

To quantify the similarity between two emotion descriptors, three complementary similarity measures are employed: Earth Mover's Distance (EMD), Jensen--Shannon (JS) similarity, and Weighted Jaccard similarity. Each metric captures a different aspect of descriptor similarity, and together they provide a comprehensive evaluation of emotional correspondence between digital objects.

\subsubsection{Earth Mover's Distance}

Earth Mover's Distance (EMD) \cite{rubner2000earth} measures the minimum transportation cost required to transform one emotional distribution into another. In the proposed framework, the transportation cost between two emotions is determined by their relative positions in the valence--arousal emotion space. Consequently, transferring emotional mass between semantically close emotions incurs a lower cost than transferring it between emotionally distant categories. The transportation cost between emotions $i$ and $j$ is defined as

\begin{equation}
c(i,j)=
\frac{
\min\left(
|\theta_i-\theta_j|,
360^\circ-|\theta_i-\theta_j|
\right)}
{180^\circ},
\end{equation}

where $\theta_i$ and $\theta_j$ denote the angular positions of the corresponding emotions. The similarity between two descriptors $P$ and $Q$ is then computed as

\begin{equation}
S_{\mathrm{EMD}}=
1-\mathrm{EMD}(P,Q).
\end{equation}

\subsubsection{Jensen--Shannon Similarity}

Jensen--Shannon similarity evaluates the similarity between two normalized emotional distributions while remaining symmetric and numerically stable \cite{lin1991divergence}. Let

\begin{equation}
M=\frac{P+Q}{2},
\end{equation}

where $P$ and $Q$ denote the aggregated emotion descriptors.

The Jensen--Shannon divergence is defined as

\begin{equation}
JS(P,Q)=
\frac{1}{2}D_{KL}(P\parallel M)+
\frac{1}{2}D_{KL}(Q\parallel M),
\end{equation}

where $D_{KL}$ denotes the Kullback--Leibler divergence.

The divergence is converted into a similarity measure as

\begin{equation}
S_{\mathrm{JS}}=
1-JS(P,Q).
\end{equation}

\subsubsection{Weighted Jaccard Similarity}

To quantify the overlap between two emotional descriptors while preserving emotion intensities, the Weighted Jaccard similarity is employed \cite{charikar2002similarity, cha2007comprehensive}:

% Unlike the classical Jaccard coefficient, which compares binary sets, the weighted formulation operates directly on continuous-valued emotional distributions:

\begin{equation}
S_{\mathrm{WJ}}(P,Q)=
\frac{
\sum_i \min(P_i,Q_i)}
{\sum_i \max(P_i,Q_i)}.
\end{equation}

This metric measures the proportion of shared emotional content between two descriptors while accounting for differences in emotion intensity.

% \colorbox{LimeGreen}{ algorithms for comparison pseudocode - Muragul}
The complete workflow for comparing two universal multimodal emotion descriptors is presented in Algorithm~\ref{alg:comparison}.

\begin{algorithm}[t]
\DontPrintSemicolon
\caption{Emotion Descriptor Comparison}
\label{alg:comparison}

\KwIn{Multimodal emotion descriptors $\mathcal{D}_1$ and $\mathcal{D}_2$}
\KwOut{Similarity scores $S_{\mathrm{EMD}}$, $S_{\mathrm{JS}}$, $S_{\mathrm{WJ}}$}

$P \leftarrow$ AggregateDescriptor$(\mathcal{D}_1)$\;

$Q \leftarrow$ AggregateDescriptor$(\mathcal{D}_2)$\;

$P \leftarrow$ Normalize$(P)$\;

$Q \leftarrow$ Normalize$(Q)$\;

$\mathbf{C} \leftarrow$ BuildCostMatrix$(\Theta)$\;

$S_{\mathrm{EMD}} \leftarrow$ EMD$(P,Q,\mathbf{C})$\;

$S_{\mathrm{JS}} \leftarrow$ JensenShannon$(P,Q)$\;

$S_{\mathrm{WJ}} \leftarrow$ WeightedJaccard$(P,Q)$\;

\Return{$S_{\mathrm{EMD}},\;S_{\mathrm{JS}},\;S_{\mathrm{WJ}}$}\;

\end{algorithm}

A higher similarity value indicates a greater correspondence between the emotional profiles of the compared artifacts, regardless of whether they originate from the same or different modalities. Consequently, paintings, musical compositions, textual documents, videos, and other multimedia content can be analyzed within a common emotional framework.

% как сравнивать два дескриптора
% формулы итд

% Given two artifacts $x$ and $y$, represented by their integrated emotional descriptors $\mathbf{V}(x)$ and $\mathbf{V}(y)$, emotional similarity can be evaluated using cosine similarity

% \begin{equation}
% \mathrm{Cosine Similarity}(x,y)
% =
% \frac{\mathbf{V}(x)\cdot\mathbf{V}(y)}
% {\|\mathbf{V}(x)\|\,\|\mathbf{V}(y)\|},
% \end{equation}

% where $\cdot$ denotes the dot product and $\|\cdot\|$ denotes the Euclidean norm.

% Alternatively, emotional dissimilarity can be measured using the Euclidean distance

% \begin{equation}
% d(x,y)
% =
% \left\|
% \right\|.
% \end{equation}

% Higher values of $\mathrm{Sim}(x,y)$ and lower values of $d(x,y)$ indicate stronger emotional correspondence between the compared artifacts.
% The proposed comparison strategy enables cross-modal emotional analysis by measuring similarity in the emotional space rather than in modality-specific feature spaces.

\section{Experiment and Results}
% \colorbox{LimeGreen}{3 example figures}

% \colorbox{Orange}{MAYBE Create fuzzy sets + fuzzy rules}

% \colorbox{LimeGreen}{12. Paragraphs must be  shortened}
\subsection{Emotion Retrieval}

\subsubsection{Image Emotion Retrieval}

Image emotion recognition is performed using the color-based approach proposed in \cite{ColorEmotionArt}. The method represents images with a fuzzy color model in the HSI color space and compares the extracted color palette with emotion-specific palettes derived from the WikiArt Dataset.

For an input image, the similarity between its color palette and each emotion palette is computed using the Jaccard coefficient:

\begin{equation}
J(E,I)=
\frac{|E \cap I|}
{|E \cup I|},
\end{equation}

where $E$ is the palette of a target emotion and $I$ is the palette extracted from the input image.

The obtained similarity values form the image emotional descriptor
 $\mathcal{D}_{img}=[d_1,d_2,\ldots,d_{10}]$ 
corresponding to the ten target emotion categories (we use seven out of them). 
 
% \begin{equation}
% \mathcal{D}_{img}=[d_1,d_2,\ldots,d_{10}],
% \end{equation}

% The overall procedure is summarized in Algorithm~\ref{alg:image_emotion}.

% \begin{algorithm}[t]
% \DontPrintSemicolon
% \caption{Image Emotion Recognition}
% \label{alg:image_emotion}

% \KwIn{Image $I$}
% \KwOut{Image emotional descriptor $\mathcal{D}_{img}$}

% $P \leftarrow$ ExtractFuzzyColorPalette$(I)$\;

% \ForEach{emotion $e_j$}{
%     $J_j \leftarrow$ JaccardSimilarity$(P, P_{e_j})$\;
% }

% $\mathcal{D}_{img}
% \leftarrow
% [J_1,J_2,\ldots,J_{10}]$\;

% \Return{$\mathcal{D}_{img}$}\;

% \end{algorithm}

\subsubsection{Lyrics Emotion Retrieval}
% \colorbox{red}{UPDATE ALGORITHM}
Lyrics are analyzed using the NRCLex library. Raw emotion scores are extracted from the input text and mapped onto a predefined emotion set used in our descriptor according to

\begin{equation}
  v_i = S(M(e_i))
\end{equation}

where \(M(e_i)\) maps the predefined emotion \(e_i\) to the corresponding NRC emotion category and \(S(\cdot)\) denotes the extracted emotion score. The resulting emotion scores are then normalized using max-normalization to obtain values in the range \([0,1]\), ensuring compatibility with descriptors extracted from other modalities.
% Lyrics are analyzed using the RoBERTa-based GoEmotions model \cite{liu2019roberta}, trained on the GoEmotions dataset \cite{demszky2020goemotions}. The model predicts confidence scores for GoEmotions categories: $G=\{g_1,g_2,\ldots,g_N\}$.

% % \begin{equation}
% % G=\{g_1,g_2,\ldots,g_N\}.
% % \end{equation}

% To obtain a unified representation across modalities, the predicted emotions are mapped to the WikiArt emotional space \cite{WikiArt} and aggregated into the proposed descriptor:

% \begin{equation}
% E_j=
% \sum_{i=1}^{N}
% w_{ij}g_i,
% \end{equation}

% where $w_{ij}$ denotes the mapping weight from the $i$-th GoEmotions category to the $j$-th target emotion. The resulting vector represents the lyrics in the proposed emotional descriptor space. 

% The complete procedure is shown in Algorithm~\ref{alg:text_emotion}.

% \begin{algorithm}[t]
% \DontPrintSemicolon
% \caption{Lyrics Emotion Recognition}
% \label{alg:text_emotion}
% \KwIn{Lyrics text $T$}
% \KwOut{Text emotional descriptor $\hat{\mathcal{D}}$}

% $G \leftarrow$ GoEmotionsRecognition$(T)$\;

% \ForEach{$g_i \in G$}{
%     $w_i \leftarrow$ WikiArtEmotionMap$(g_i)$\;
% }

% \ForEach{$w_i \in W$}{
%     $d_i \leftarrow$ DescriptorMap$(w_i)$\;
% }

% $\hat{\mathcal{D}} \leftarrow$ Normalize$(D)$\;

% \Return{$\hat{\mathcal{D}}$}\;

% \end{algorithm}

\begin{figure*}[ht!]
    \centering

    \begin{subfigure}[t]{0.45\textwidth}
        \centering
        \includegraphics[width=\linewidth]{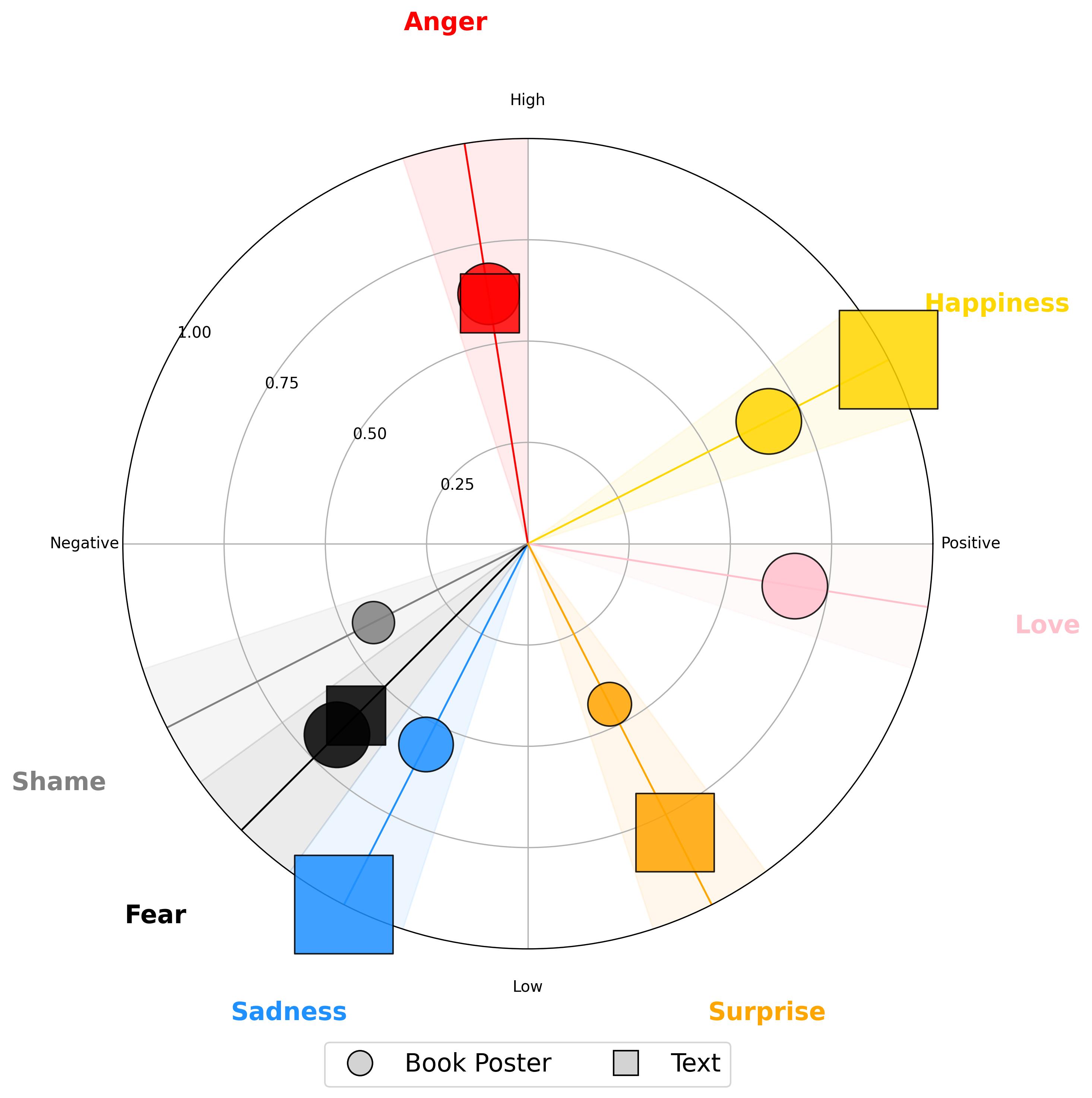}
        \caption{Book descriptor}
        \label{fig:mbu_book}
    \end{subfigure}
    \hfill
    \begin{subfigure}[t]{0.45\textwidth}
        \centering
        \includegraphics[width=\linewidth]{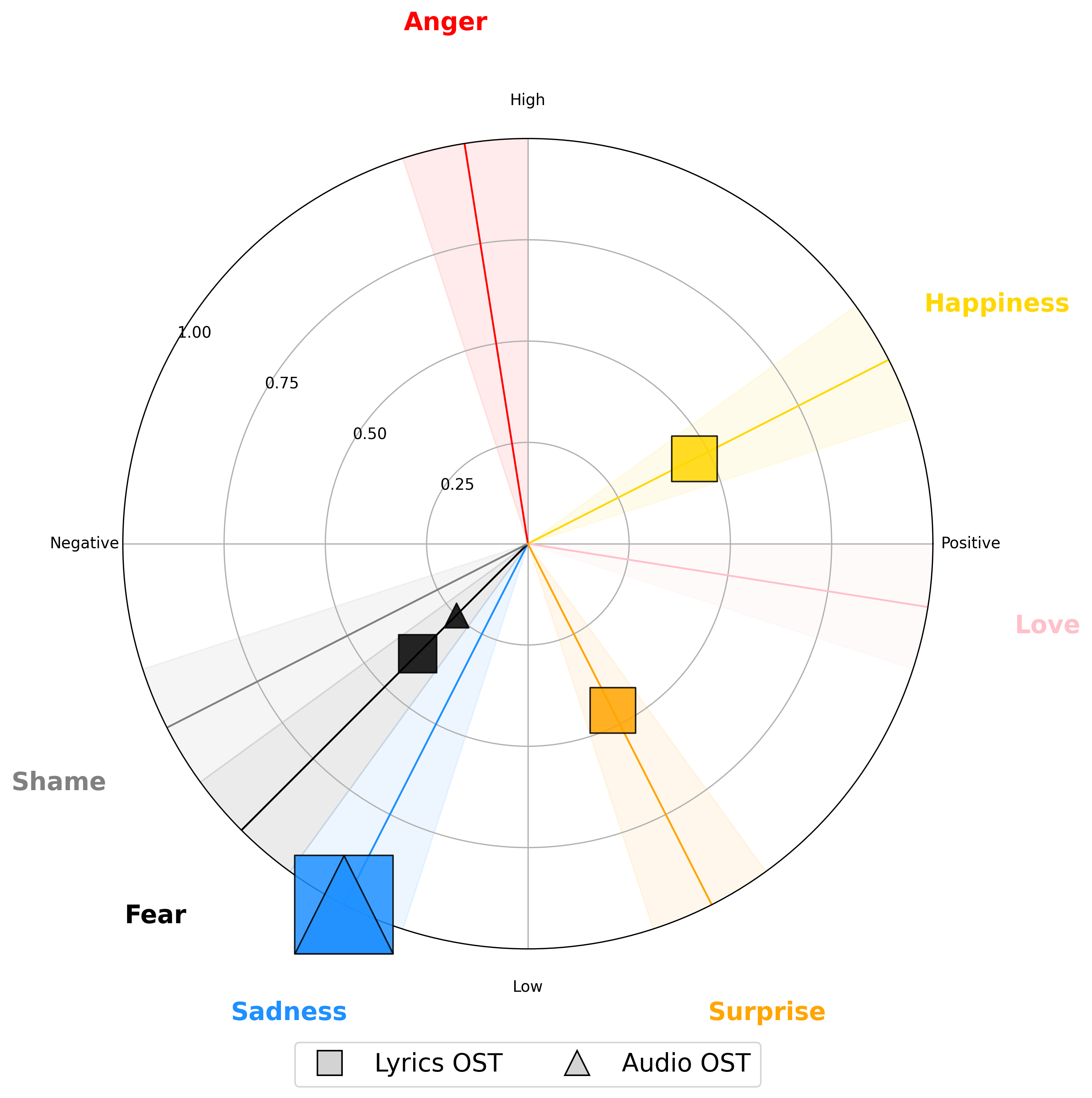}
        \caption{OST descriptor}
        \label{fig:mbu_ost}
    \end{subfigure}

    \caption{Comparison of emotional descriptors for the book and OST of \textit{Me Before You}.}
    \label{fig:mbu}
\end{figure*}

\begin{figure*}[ht!]
    \centering

    \begin{subfigure}[t]{0.45\textwidth}
        \centering
        \includegraphics[width=\linewidth]{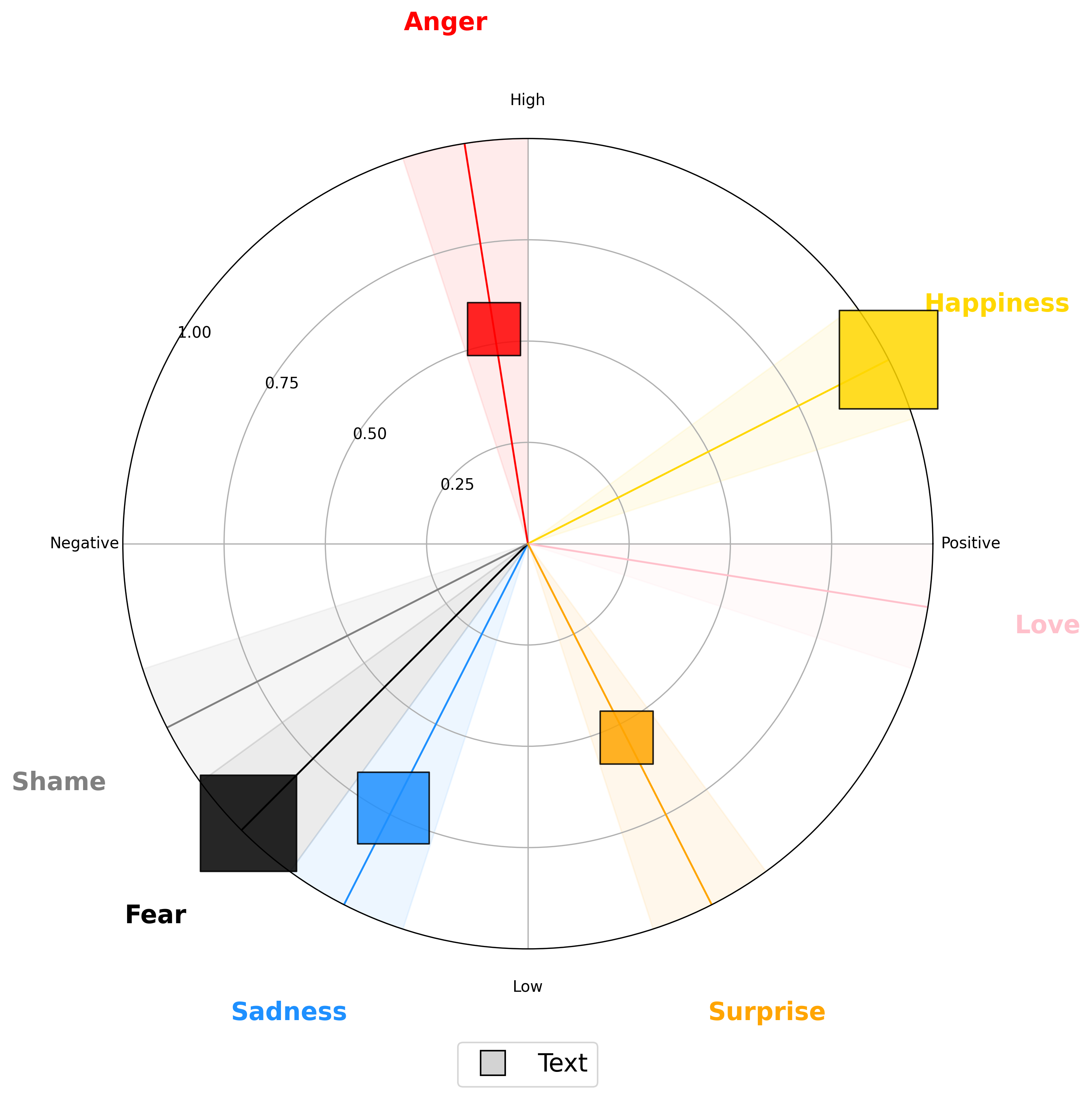}
        \caption{Book descriptor}
    \end{subfigure}
    \hfill
    \begin{subfigure}[t]{0.45\textwidth}
        \centering
        \includegraphics[width=\linewidth]{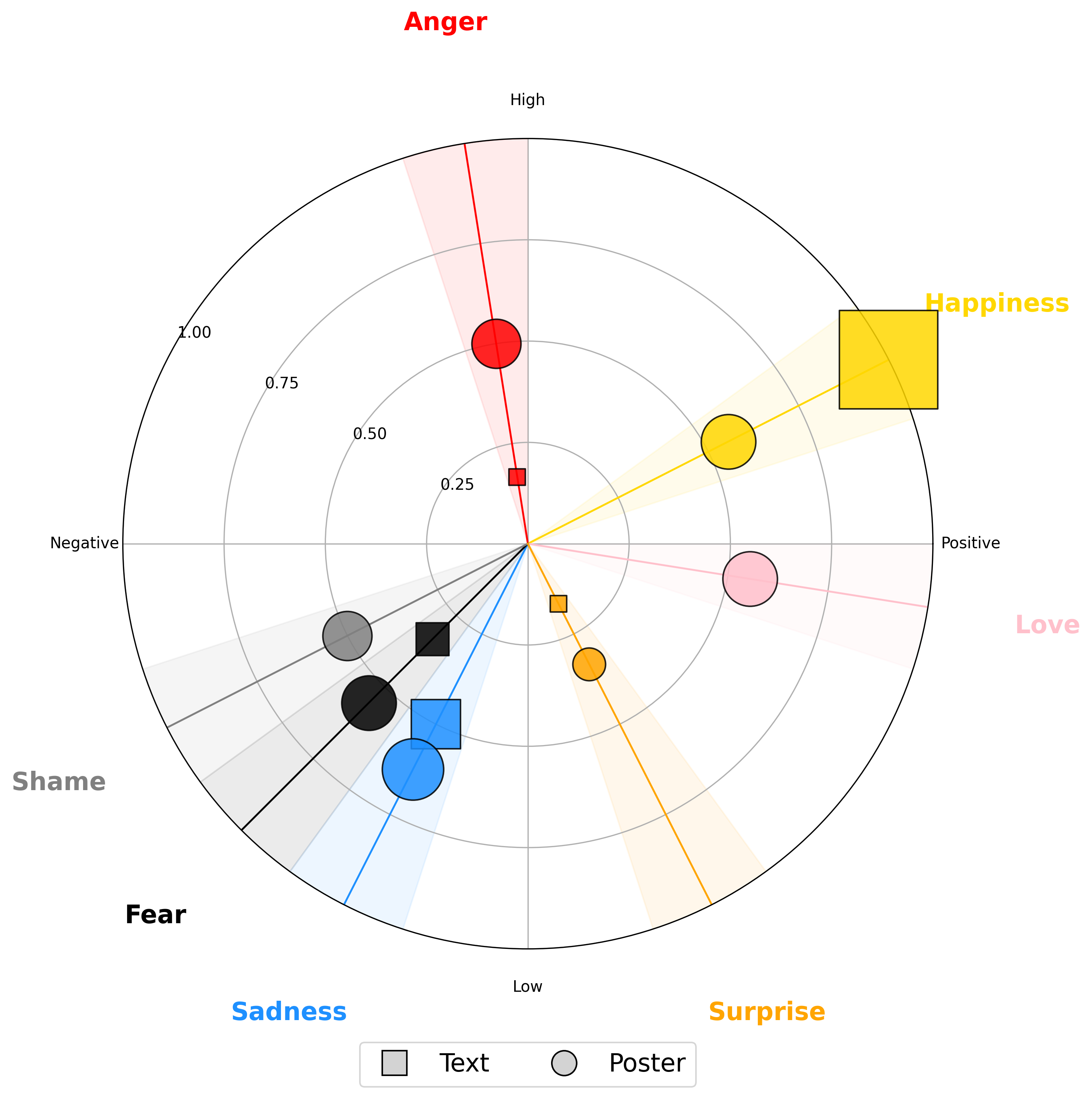}
        \caption{Movie descriptor}
    \end{subfigure}

    \caption{Comparison of emotional descriptors for \textit{Anna Karenina} and its 2012 film adaptation.}
\end{figure*}

\begin{figure*}[ht!]
    \centering

    \begin{subfigure}[t]{0.45\textwidth}
        \centering
        \includegraphics[width=\linewidth]{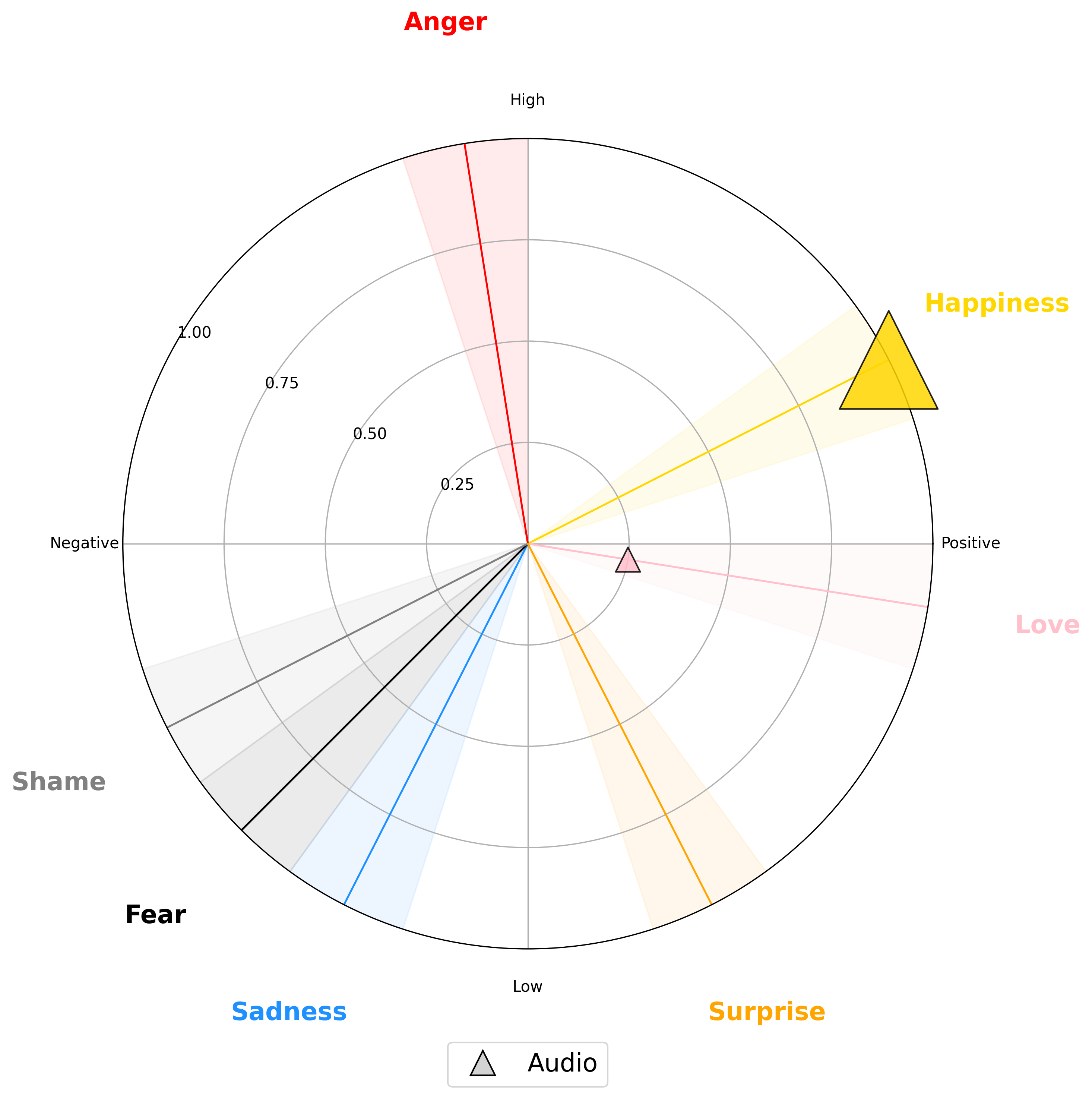}
        \caption{\textit{Happy} by Pharrell Williams}
    \end{subfigure}
    \hfill
    \begin{subfigure}[t]{0.45\textwidth}
        \centering
        \includegraphics[width=\linewidth]{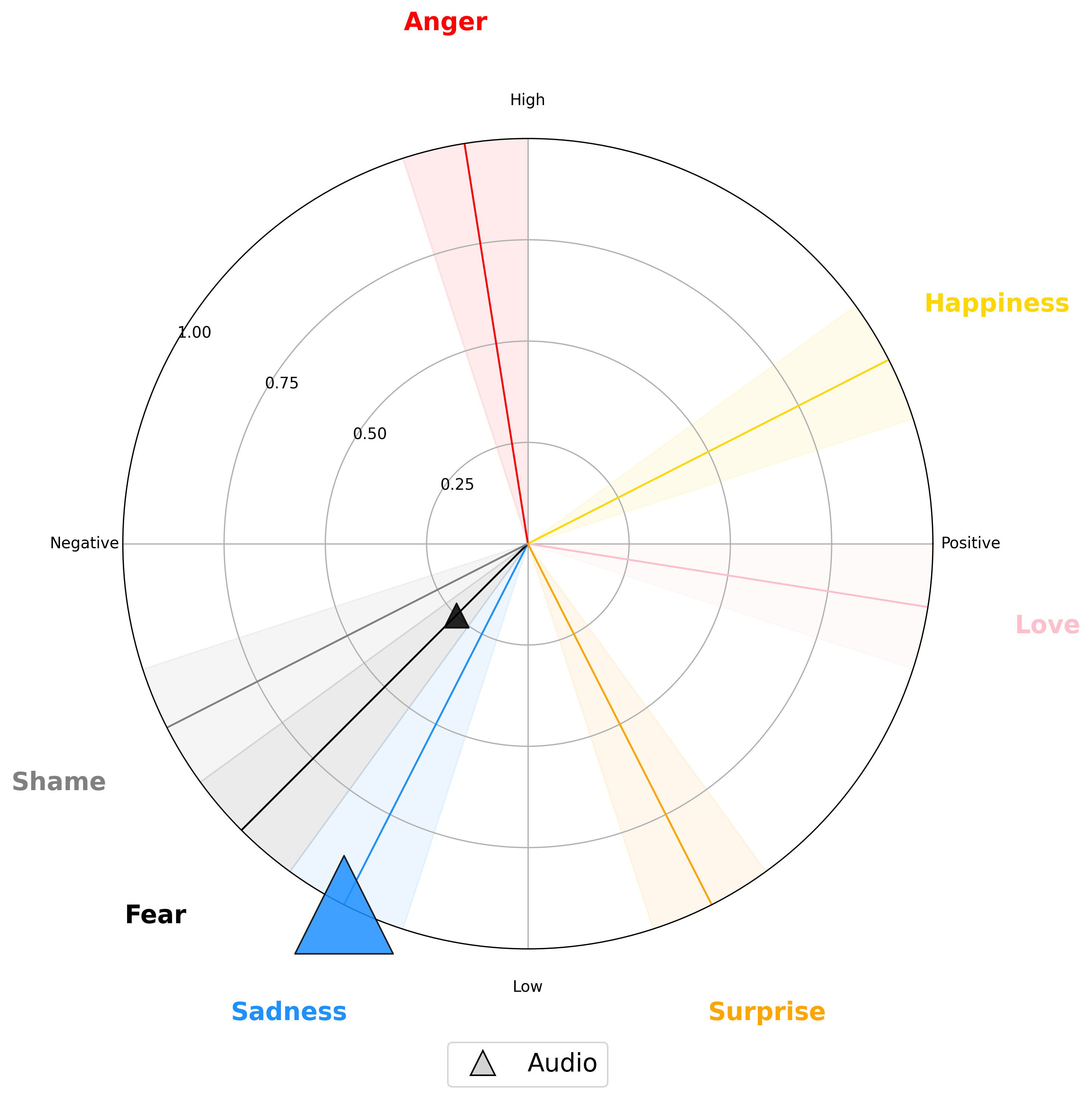}
        \caption{\textit{Another Love} by Tom Odell}
    \end{subfigure}

    \caption{Comparison of emotional descriptors for two songs with contrasting emotional profiles.}
\end{figure*}

\begin{table*}[ht!]
\centering
\caption{Similarity scores between universal multimodal emotion descriptors computed using Earth Mover's Distance (EMD), Jensen--Shannon (JS), and Weighted Jaccard (WJ) similarity.}
\label{tab:experiment_results}

% \begin{tabular}{|p{3.4cm}|p{3.4cm}|c|c|c|}
\begin{tabular}{|l|l|c|c|c|}
\hline
\textbf{Object 1} &
\textbf{Object 2} &
\textbf{EMD} &
\textbf{JS} &
\textbf{WJ} \\
\hline

\textit{Me Before You} OST (Lyrics + Audio)
&
\textit{Me Before You} Book (Poster + Text)
&
0.8311 &
0.6255 &
0.5046 \\
\hline

\textit{Anna Karenina} Book (Text)
&
\textit{Anna Karenina} Movie 2012 (Poster + Text)
&
0.8840 &
0.5579 &
0.3761 \\
\hline

Happy (Audio)
&
Another Love (Audio)
&
0.2200 &
0.1674 &
0.0000 \\
\hline

\end{tabular}
\end{table*}

\subsubsection{Audio Emotion Retrieval}

% \colorbox{LimeGreen}{Add reference to bibtex PMEmo2019}
Music emotion retrieval is based on 30-second audio excerpts obtained via the Deezer API. Each excerpt is divided into 5 equal segments, converted into Mel-spectrograms, and processed by an emotion regression model trained on the PMEmo dataset \cite{PMEmo2018}. For each segment, the model predicts valence and arousal values, which are subsequently mapped into the proposed emotional descriptor space:

% \begin{equation}
% (v_k,r_k)=\Phi(M_k),
% \end{equation}

% \begin{equation}
% d_k=DescriptorMap(v_k,r_k).
% \end{equation}

The resulting $\mathcal{D}_{audio}=\{d_1,d_2,\ldots,d_K\}$ preserves the temporal evolution of emotions and serves as the audio emotional descriptor. 
% \begin{equation}
% \hat{\mathcal{D}}^{audio}
% =
% \{d_1,d_2,\ldots,d_K\}
% \end{equation}

% Algorithm~\ref{alg:audio_emotion} summarizes the retrieval process.

% \begin{algorithm}[t]
% \DontPrintSemicolon
% \caption{Audio Emotion Retrieval}
% \label{alg:audio_emotion}

% \KwIn{Audio excerpt $A$}
% \KwOut{Audio emotional trajectory $\hat{\mathcal{D}}^{audio}$}

% $S \leftarrow$ Split$(A,3s)$;

% \ForEach{$s_k \in S$}{

% $M_k \leftarrow$ ExtractMelSpectrogram$(s_k)$;

% $(v_k,r_k) \leftarrow$ PMEmoModel$(M_k)$\;

% $d_k \leftarrow$ DescriptorMap$(v_k,r_k)$\;

% }

% $\hat{\mathcal{D}}^{audio}\leftarrow\{d_1,d_2,\ldots,d_K\}$;

% \Return{$\hat{\mathcal{D}}^{audio}$};

% \end{algorithm}

\subsection{Comparison}
% \paragraph{Book Emotion Descriptor}
% \begin{figure}
%     \centering
%     \includegraphics[width=\linewidth]{fig/example1.jpeg}
%     \caption{Enter Caption}
%     \label{fig:placeholder}
% \end{figure}

% \paragraph{Movie Emotion Descriptor}
% \begin{figure}
%     \centering
%     \includegraphics[width=\linewidth]{fig/example2.jpeg}
%     \caption{Enter Caption}
%     \label{fig:placeholder}
% \end{figure}

% \subsubsection{Comparison and Evaluation}

% \colorbox{LimeGreen}{Table with comparison results}
To provide an illustrative evaluation of the proposed universal multimodal emotion descriptor, three representative pairs of digital objects were selected. The first two pairs correspond to semantically related content originating from the same creative work but represented by different modalities: the Me Before You original soundtrack (lyrics and audio) compared with the corresponding book (poster and a textual summary of the book), and Leo Tolstoy's novel Anna Karenina compared with its 2012 film adaptation, where the textual representation of the novel is based on its summary. As a contrasting example, a pair consisting of a happy song (Pharrell Williams - \textit{Happy}) and a sad song (Tom Odell - \textit{Another Love}) was included to evaluate the proposed descriptor's ability to distinguish emotionally dissimilar content.

% \section{Results}
The similarity between the aggregated emotion descriptors was evaluated using Earth Mover's Distance (EMD), Jensen--Shannon (JS) similarity, and Weighted Jaccard (WJ) similarity. The obtained similarity scores are summarized in Table~\ref{tab:experiment_results}.

As expected, both semantically related pairs achieved considerably higher similarity scores than the emotionally contrasting pair across all three evaluation metrics. In particular, the comparison between the \textit{Anna Karenina} book and its film adaptation yielded the highest EMD similarity ($0.8840$), indicating a strong correspondence in their emotional representations despite being derived from different modalities. Similarly, the \textit{Me Before You} soundtrack and the corresponding book exhibited consistently high similarity values. In contrast, the comparison between the happy and sad songs produced substantially lower similarity scores, with the Weighted Jaccard similarity equal to zero, indicating no overlap between their dominant emotional distributions. These observations demonstrate that the proposed descriptor preserves emotional similarity across heterogeneous modalities while effectively distinguishing emotionally contrasting digital objects.

% \colorbox{LimeGreen}{Describe Table with comparison results}

\section{Conclusion}
This paper presented CD-MED, a Cross-Domain Multimodal Emotion Descriptor for representing heterogeneous digital objects within a unified visual emotional space. By transforming modality-specific emotion recognition outputs into a common emotional representation, the proposed approach enables direct visual comparison of digital objects regardless of their modality composition. The descriptor provides both an interpretable visualization in the valence--arousal space and a normalized emotional representation suitable for similarity analysis. Experimental results demonstrated that semantically related objects consistently achieved higher similarity scores than emotionally contrasting objects using Earth Mover's Distance, Jensen--Shannon similarity, and Weighted Jaccard similarity.

% \colorbox{LimeGreen}{13. Discussion}
% The proposed framework demonstrates that heterogeneous multimedia content can be representein a shared emotional space without requiring a unified emotion-recognitionon model. Instead, modality-specific models are treated as interchangeable components whose outputs are projected into a common descriptor. 

The framework is flexible and easily extensible while preserving interpretability through a unified visual representation. At the same time, the current evaluation is intended as a proof of concept and is limited to a small number of representative examples. A more comprehensive quantitative evaluation is required to further validate the robustness and generalization capability of the proposed descriptor. 

% Another promising direction is the development of descriptor-specific similarity measures that further exploit the structure of the proposed emotional representation.

% % \colorbox{LimeGreen}{14. Future works}
% Future work will focus on extending the descriptor to additional modalities, including video and physiological signals, investigating alternative aggregation strategies for multimodal emotional evidence, and conducting large-scale evaluations on publicly available multimedia datasets. 

\section*{Acknowledgement}
This research has been funded by the Science Committee of the Ministry of Science and Higher Education of the Republic of Kazakhstan (Grant No. AP22786412)

\bibliography{export}

\end{document}